\documentclass[ onecolumn,amsmath,amssymb,12pt,superscriptaddress,nofootinbib]{revtex4}
\pdfoutput=1
\usepackage{float}
\usepackage[latin1]{inputenc}
\usepackage[english]{babel}
\usepackage[normalem]{ulem}
\usepackage{amssymb}
\usepackage{amsmath}
\usepackage{comment}
\usepackage{amsthm}
\usepackage[]{graphicx}
\usepackage{tensor}
\usepackage{color}
\usepackage{cancel}
\usepackage{setspace}
\usepackage{fancyhdr}
\usepackage{framed}
\usepackage{tikz}
\usepackage[bookmarks,linktocpage, colorlinks=true, plainpages = false, citecolor = blue,  linkcolor=blue, urlcolor = blue, filecolor = blue]{hyperref} 
\usepackage{subcaption}
\usepackage{natbib}
\graphicspath{{./figs/}}
\usepackage[titles]{tocloft}

\newcommand{\new}[1]{#1}

\begin{document}

\allowdisplaybreaks
\begin{titlepage}

\title{No-boundary prescriptions in Lorentzian quantum cosmology}

\author{Alice Di Tucci}
\email{alice.di-tucci@aei.mpg.de}
\affiliation{Max--Planck--Institute for Gravitational Physics (Albert--Einstein--Institute), 14476 Potsdam, Germany}
\author{Jean-Luc Lehners}
\email{jlehners@aei.mpg.de}
\affiliation{Max--Planck--Institute for Gravitational Physics (Albert--Einstein--Institute), 14476 Potsdam, Germany}
\author{Laura Sberna}
\email{lsberna@perimeterinstitute.ca}
\affiliation{Perimeter Institute, 31 Caroline St N, Ontario, Canada}

\begin{abstract}
\vspace{.3in} \noindent We analyse the impact of various boundary conditions on the (minisuperspace) Lorentzian gravitational path integral. In particular we assess the implications for the Hartle-Hawking no-boundary wavefunction. It was shown recently that when this proposal is defined as a sum over compact metrics, problems arise with the stability of fluctuations. These difficulties can be overcome by an especially simple implementation of the no-boundary idea: namely to take the Einstein-Hilbert action at face value while adding \emph{no boundary term}. This prescription simultaneously imposes an initial Neumann boundary condition for the scale factor of the universe and, for a Bianchi IX spacetime,  Dirichlet conditions for the anisotropies. Another way to implement the no-boundary wavefunction is to use Robin boundary conditions. A sub-class of Robin conditions allows one to specify the Hubble rate on the boundary hypersurface, and we highlight the surprising aspect that specifying the final Hubble rate (rather than the final size of the universe) significantly alters the off-shell structure of the path integral.  The conclusion of our investigations is that all current working examples of the no-boundary wavefunction force one to abandon the notion of a sum over compact and regular geometries, and point to the importance of an initial Euclidean momentum.
\end{abstract}

\maketitle

\end{titlepage}

\tableofcontents

\section{Introduction}	

Quantum theory is based on calculating transition amplitudes. In particular, when a system is prepared in a certain state, we can ask: what is the probability for various outcomes? Quantum cosmology applies this framework to the universe. In that case, the question would be: if the universe is in a certain state at a certain time, what is the probability for it to evolve to different later configurations? Asking such questions presupposes that we know the state of the universe at a certain time. But as we extrapolate our knowledge of the universe back into the more distant past, we know less and less about the state of the universe. Yet, everything followed from these early conditions. 

When gravity is involved, transition amplitudes are calculated between 3-geometries, which in the cosmological context may usually be thought of as equal-time slices of the universe's evolution. Instead of having to specify conditions on ever earlier such initial slices, Hartle and Hawking had the beautiful idea that one could calculate transition amplitudes which have no boundary in the past, i.e. transitions amplitudes involving only a specification of the late time configuration and no initial 3-geometry \cite{Hartle:1983ai}. The idea was that this proposal could describe how the universe originated from nothing, i.e. how spacetime and matter arose from the absence thereof. Moreover, this proposal would implicitly fix the initial conditions of the universe \cite{Hartle:2008ng}.   

A question which has vexed quantum cosmologists since the appearance of this proposal has however been how to actually calculate such no-boundary amplitudes in practice. Transition amplitudes are naturally expressed as gravitational path integrals, in an extension of Feynman's path integral quantisation programme to include gravity. Hartle and Hawking proposed that in this framework one should sum only over geometries that are compact and regular in the past, in order to implement their idea. Yet, it still remains difficult to perform such a calculation in practice. Even for simple examples in minisuperspace it has remained unclear how to precisely define the various required boundary conditions, integration ranges and integration contours (for an early investigation see \cite{Halliwell:1988ik}). A priori, it sounds like the absence of a boundary would eliminate the need for boundary conditions. But in fact one basic difficulty, which arises as a direct consequence of the $1+3$ split of spacetime, is that for each field that is considered one must impose \emph{some} conditions at the end points of integration, i.e. one must decide which boundary conditions best describe the absence of a boundary.

In the present paper we will study the consequences of imposing various boundary conditions, both at the initial ``no-boundary'' hypersurface, and also at late times. In fact, one surprising aspect of our work is that changing the late time boundary conditions can significantly affect the path integral over its entire range of integration. In terms of choices of contour, we will stick as much as possible to the most conservative choice, which is to integrate over real field configurations, and thus Lorentzian geometries\footnote{The integrals that we will consider typically do not contain a singularity at the origin $N=0$ of the complexified plane of the lapse function $N,$ and hence we will mainly be concerned with integrations over infinite ranges of the lapse. This means that our results will mostly pertain to wavefunctions, and not to Green's functions.}. In other words, we will not attempt to define a Euclidean gravitational path integral, which in most cases is ill-defined and divergent in any case \cite{FLT1}. We will however encounter one example where we will be forced to depart from the strictly Lorentzian integral.

Overall, we will take the point of view that the no-boundary wavefunction is a result in need of a definition. Thus, we will take the Hartle-Hawking saddle point geometries, which we will review in section \ref{sec:noboundaryterm}, as the basic building blocks and we will seek a well-defined path integral that has these geometries (and ideally only these) as its relevant saddle points. Prior works have studied various potential definitions mainly by exploring different contours of integration for the lapse integral, see e.g. \cite{Halliwell:1988ik, Halliwell:1990tu, DiazDorronsoro:2017hti, Feldbrugge:2017mbc}. Here we will try an alternative route, which is to explore different possibilities for the boundary conditions \cite{Halliwell:1988ik, DiazDorronsoro:2018wro, DiTucci:2019dji}. Our starting point is the most straightforward set of boundary conditions, namely Dirichlet boundary conditions. The Dirichlet condition can be used to set the size of the universe to zero on the initial hypersurface, and can hence directly implement the idea of summing over compact geometries. However, these boundary conditions immediately lead to problems with the stability of perturbations \cite{FLT2}. In light of this, we will explore the consequences of Neumann conditions that fix the initial momentum, and of Robin conditions that fix a combination of both field value and momentum. Robin conditions seem eminently useful and physical, as they allow one to specify the Hubble rate at early or late times, for example. These various types of boundary condition will first be reviewed in section \ref{sec:generalsetup}, and then applied to the present context in the rest of the paper. Our paper thus extends and generalises the work begun in \cite{DiTucci:2019dji}. We will find that the specification of an (Euclidean) initial momentum is rather crucial to the success of the programme, and we will discuss three possible implementations of the no-boundary wavefunction as a path integral. Each of these has some advantages and drawbacks, which we will discuss, and which may eventually reduce the number of viable proposals. Our conclusions and a list of future directions will be provided in section \ref{sec:discussion}.


\section{General setup}\label{sec:generalsetup}

\begin{figure}
	\centering
	\includegraphics[width=0.6\textwidth]{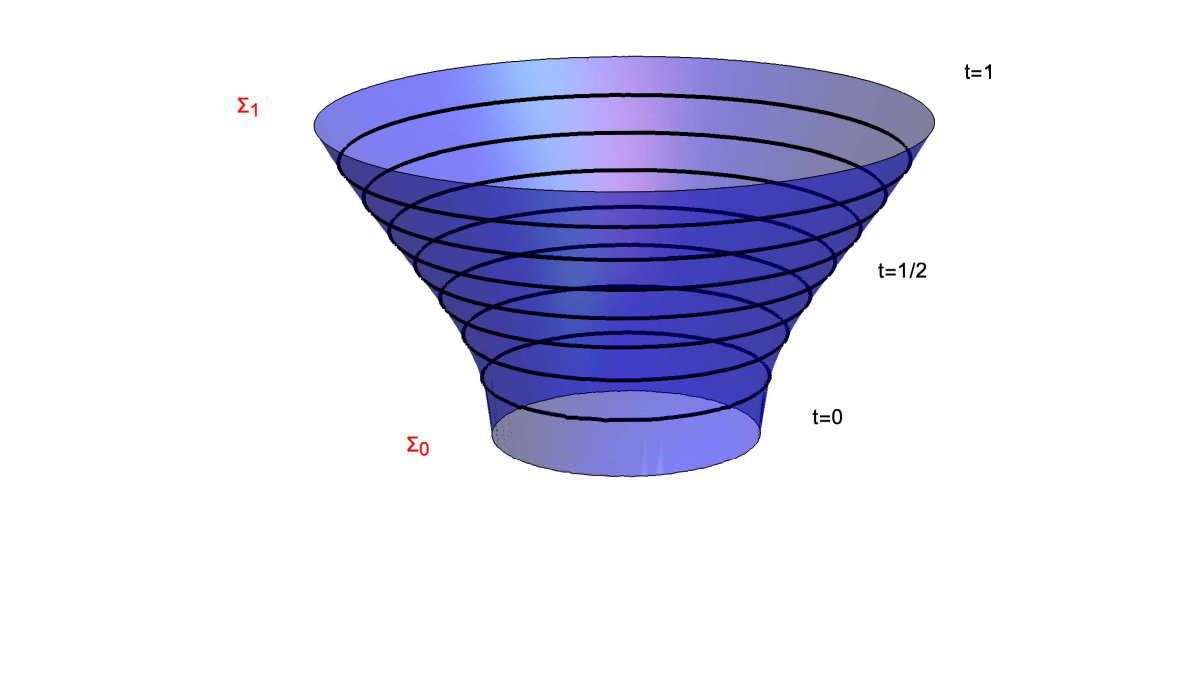} 
	\caption{A transition amplitude in quantum cosmology: we sum over all metrics interpolating between two hypersurfaces $\Sigma_{0,1}$ located at coordinate times $t=0,1$ and obeying specific boundary conditions on $\Sigma_{0,1}$. The present paper explores boundary conditions that allow for an implementation of the no-boundary wavefunction.
	}
	\label{fig:amplitude}
\end{figure}

We start with the Einstein-Hilbert action in the presence of a cosmological constant $\Lambda \equiv 3 H^2$,
\begin{equation}\label{eq:EH}
S_{\text{EH}}= \frac{1}{2} \int_M d^4 x \,  \sqrt{-g}  \, (R - 6 H^2) \, ,
\end{equation}
where we have set $8\pi G =1.$ We will be interested in evaluating the path integral
\begin{equation}
\Psi = \int_{\Sigma_0}^{\Sigma_1} dN \, \delta q \, e^{i S/\hbar}
\end{equation}
in a minisuperspace context, see Fig. \ref{fig:amplitude}. This means that we will only consider finitely many free functions in the metric. In the simplest case we will restrict to just the scale factor $a$ of the universe and the lapse function $N$. Then a convenient form of the metric is \cite{Halliwell:1988wc} 
\begin{equation}\label{eq:ansatz}
ds^2  = - \frac{N^2 }{q} dt^2 + q \, d  \Omega_3^2,
\end{equation}
where $q=a^2$ is the scale factor squared and the lapse $N$ determines the total time elapsed between the initial hypersurface $\Sigma_0$ at $t=0$ and the final hypersurface $\Sigma_1$ at coordinate time $t=1$\new{, according to $ t_{\rm{phys}} = \int_{0}^{1} dt \, N \, |q|^{-1/2} $ (in cases where the geometry contains a complex/Euclidean part, one might have to restrict the integration range to the part of the geometry that is real and Lorentzian, in order to obtain a quantity that one can interpret as the actual physical time)}. Here we have taken the spatial part of the metric to be given by the metric on a 3-sphere $d\Omega_3^2$ with volume $V_3.$ The utility of \eqref{eq:ansatz} stems from the fact that the kinetic term for the scale factor becomes quadratic in $q$,
\begin{equation}\label{eq:EHansatz}
\frac{S}{V_3} = \frac{S_{\text{EH}}}{V_3} +S_B= \int_0^1 dt \, \Bigl[ \frac{3}{2 N} q \, \ddot{q} + \frac{3}{4 N} \dot{q}^2 + 3 N (1 - H^2 q) \Bigr] +S_B\,,
\end{equation}
where we added possible boundary contributions $S_B$ localised on the hypersurfaces at $t=0,1.$ The boundary contributions will be necessary to obtain a consistent variational problem, as can be seen by varying the action w.r.t. $q,$ (note that we can work in a gauge where $N$ is fixed while performing the variation w.r.t. $q$, see \cite{Teitelboim:1981ua, Teitelboim:1983fk, Halliwell:1988wc} for details)
\begin{align}
\frac{\delta_q S}{V_3} = \int_0^1 dt \, \Bigl[\frac{3}{2N}\ddot{q} -3NH^2\Bigr] \delta q  + \frac{3}{2N}q\, \delta\dot{q}\mid^{t=1}_{t=0} + \delta S_B\,. \label{variation}
\end{align}
In this manner we obtain the equation of motion for the scale factor, 
\begin{align}
\ddot{q} = 2N^2H^2\,, \label{eomq}
\end{align} 
which must be solved subject to appropriate boundary conditions at $t=0,1.$ The following are the usual options:
\begin{itemize}
\item \emph{Neumann} boundary conditions, $S_B=0$~\cite{Krishnan:2016mcj}.
If we do not add any boundary term, then Eq. \eqref{variation} shows that the scale factor automatically inherits a Neumann boundary condition at $t=0,1,$ 
\begin{align}
\frac{3}{2N}q\,\delta\dot{q}\mid^{t=1}_{t=0} = 0\,, \label{Neumann}
\end{align}
implying that we can set the momentum $\dot{q}/N$ to any desired value at the end points, without fixing the field value itself.
\item \emph{Dirichlet} boundary conditions, $S_B=-\frac{3}{2N}q\dot{q}.$ In this case the variation of the boundary term removes the term in $\delta \dot{q}$ and replaces it with a term in $\delta q$ only,
\begin{align}
-\frac{3}{2N}\dot{q}\,\delta q \mid^{t=1}_{t=0} = 0\,, \label{Dirichlet}
\end{align}
implying that we can set the field value $q$ to any desired value at the end points, without fixing the momentum. The required boundary term is the well-known Gibbons-Hawking-York (GHY) term \cite{York:1972sj, Gibbons:1976ue} written out for our metric, $\int d^3x \sqrt{h}K=-\frac{3}{2N}q\dot{q}\mid_{bdy}$ where $h$ denotes the determinant of the induced metric on the boundary and $K$ is the trace of the extrinsic curvature. 
\item \emph{Robin} boundary conditions, $S_B = f(q).$ In this case the variation of the boundary action w.r.t. $q$ yields a condition on a combination of both the field value and the momentum, since the variation gives
\begin{align}
\frac{3}{2N} q \,\delta \dot{q} + f_{,q} \delta q \mid^{t=1}_{t=0}\, = \frac{3q}{2} \,\delta\left(\frac{1}{N}\dot{q} + \frac{2}{3} g(q) \right)\mid^{t=1}_{t=0} \qquad \text{where}\quad g(q) = \int \frac{f_{,q}}{q} dq\,, \label{Robin}
\end{align}
so that the combination $\frac{1}{N}\dot{q} + \frac{2}{3} g(q)$ can be set to any desired value at the end points. Note that a Robin boundary condition is in no way exotic: when we specify the Hubble rate on a given hypersurface, we are in effect imposing a Robin condition: say we specify $H = \frac{a_{,t_{p}}}{a} \equiv {\cal H},$ then we can rewrite this condition in Robin form as $a_{,t_{p}} - {\cal H} a = 0,$ where $t_{p}$ denotes the physical time. \new{In full gravity, a standard choice for a Robin condition would be $ \delta \left(\pi^{ij} + \xi \sqrt{|h|} h^{ij} \right)=0$, where $ h_{ij} $ is the metric of the boundary and $ \pi^{ij} $ its conjugate momentum~\cite{Krishnan:2017bte}.}
\item \emph{Special} boundary conditions arise when the prefactor of the variation is set to zero. For instance, we may obtain a Special Neumann condition if in \eqref{Dirichlet}, instead of setting $\delta q=0,$ we set $\dot{q}=0$ at the boundary. The variational problem is then also well-defined, but note that this only works for the special case where the momentum is set to zero. If one wants to set the momentum to a non-zero value, one must use the Neumann condition \eqref{Neumann} instead, and this requires a different boundary term. In a similar vein, one may set $q=0$ on the boundary thereby converting \eqref{Neumann} into a Special Dirichlet condition. We will encounter slightly more general examples of such Special boundary conditions in section \ref{sec:canonical}, where we will implement a Special Robin condition.
\end{itemize}
\new{Dirichlet and Neumann are widely known and adequate conditions under most circumstances. While less common, mixed (Robin) conditions have already proven useful for some gravitational problems, e.g. the formal definition of perturbation theory in Euclidean gravity~\cite{Witten:2018lgb}.}
In the context of general relativity, an important question regarding all such boundary terms is whether or not they can be given a covariant expression \cite{York:1972sj, Gibbons:1976ue, Krishnan:2017bte}. We will discuss this further in the coming sections. 

Once we have specified a set of boundary conditions, we can proceed with the evaluation of the path integral. The general procedure that we employ here exploits the fact that the action is quadratic in $q.$ As first discussed in this context in \cite{Halliwell:1988ik} (and reviewed in \cite{FLT1}), one may shift the variable of integration to $Q$ defined as $q(t) \equiv \bar{q}(t) + Q(t)$ where $\bar{q}$ denotes a solution of the equation of motion \eqref{eomq} for $q$ with the specified boundary conditions, and $Q$ is a perturbation which is completely general except that it must respect the boundary conditions at $t=0,1$. The integral over $q$ then splits into the integrand evaluated over the classical solution $ \bar{q} $, and a Gaussian integral over the perturbations $Q.$ This last integral can also be performed in a standard fashion, and simply affects the prefactor of the path integral. Since we will ultimately only be interested in the leading order saddle point approximation to the path integral, we will neglect such prefactors.

Once these integrations have been performed, we are left with an ordinary integral over the lapse function $N.$ In general, this is a complicated oscillatory integral that cannot be calculated exactly. In fact, given that the integral is typically only conditionally convergent, one must take care in defining and evaluating it. The appropriate mathematical framework for evaluating such an integral is Picard-Lefschetz theory (see \cite{Witten:2010cx,FLT1}), which can be thought of as a way of using Cauchy's theorem in a systematically efficient manner. Specific examples of boundary conditions, to which we now turn, will serve to illustrate the general procedure that we have just outlined.


\section{The simplest case: the ``No Boundary Term'' proposal} \label{sec:noboundaryterm}

\begin{figure}
	\centering
	\includegraphics[width=0.4\textwidth]{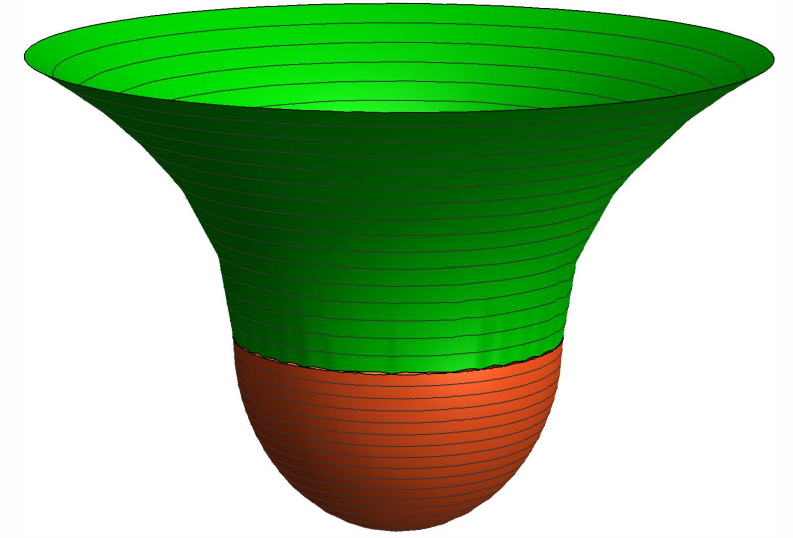}
	\includegraphics[width=0.4\textwidth]{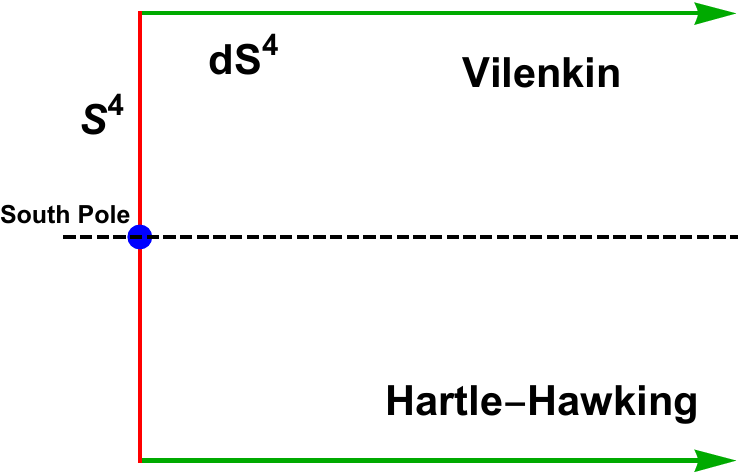} 
	\caption{{\it Left panel:} The Hartle-Hawking/Vilenkin geometry, typically represented as a Lorentzian geometry (in green) glued onto a Euclidean geometry (in red). {\it Right panel:}  Green lines with arrowheads indicate directions of increasing real/Lorentzian time, red lines depict imaginary/Euclidean time directions. The Euclidean extensions of the geometries imply specific Wick rotations.
	}
	\label{fig:Wick}
\end{figure}

The Hartle-Hawking (HH) wavefunction \cite{Hawking:1981gb, Hartle:1983ai} is best known in terms of the iconic geometry depicted on the left in Fig. \ref{fig:Wick}. In fact, the same picture would apply to the closely related tunneling geometry explored by Vilenkin (V) \cite{Vilenkin:1982de}. It will be useful to briefly review these geometries. We may start with the de Sitter geometry, which provides a solution to Einstein's equations in the presence of a positive cosmological constant $\Lambda = 3 H^2,$ 
\begin{align}
ds^2 = - N_p^2 dt_p^2 + \frac{1}{H^2}\cosh^2(Ht_p) d\Omega_3^2\,,
\end{align}
where $d\Omega_3^2$ represents the metric on a 3-sphere and the lapse is simply set to $N_p=1$. When embedded in 5 dimensions, de Sitter space in the closed slicing can be pictured as a hyperboloid with minimum spatial extent at $t_p=0.$ The intuition behind the no-boundary wavefunction is that the geometry should be rounded off, so as to have no boundary. This can be achieved by analytically continuing the solution to Euclidean time, starting precisely from the waist of the hyperboloid at $t_p=0.$  Thus one may set
\begin{align}
t_p= \mp i \left(\tau - \frac{\pi}{2H}\right)\,, \qquad \frac{\pi}{2H} \geq \tau \geq 0 \label{ancon}
\end{align}
with the result that along the Euclidean section the metric turns into that of a 4-sphere
\begin{align}
ds^2 = d\tau^2 + \frac{1}{H^2}\sin^2(H\tau) d\Omega_3^2\,.
\end{align}
The geometry then smoothly closes off at $\tau=0,$ a coordinate location sometimes referred to as the South Pole. 

We can see that there are two possibilities for the rotation to Euclidean time, corresponding to the two different signs in \eqref{ancon}\new{\footnote{Mathematically changing $i$ to $-i$ is an isomorphism, and should not change physical quantities. But in the physics that we discuss here there are in fact two factors of $i$: one comes from the definition of the path integral, which is weighted by $e^{iS}$. The second one arises when the action $S$ itself picks up an imaginary part, as is the case for no-boundary and tunneling geometries. Thus the relative sign between no-boundary and tunneling geometries indeed matters, as it leads to an inverse weighting when evaluating $e^{iS}$. This weighting is independent of the isomorphism.}}. 
These choices correspond to two different Wick rotations. The upper sign coincides with the usual Wick rotation applied in quantum field theory -- this is the choice made by Hartle and Hawking and it has the consequence that perturbations around this geometry are stable and suppressed \cite{Hartle:1983ai, Halliwell:1984eu}. The action of the Euclidean section of the HH geometry is given by $-i\frac{4\pi^2}{H^2}.$ The lower choice of sign in \eqref{ancon} yields Vilenkin's tunneling geometry. The action along its Euclidean section is given by $+i\frac{4\pi^2}{H^2},$ while small perturbations around this geometry are unsuppressed \cite{Halliwell:1989dy, FLT2}\footnote{In their recent papers Vilenkin and Yamada \cite{Vilenkin:2018dch, Vilenkin:2018oja} dispute this conclusion.}. Note that one may also think of the continuation to Euclidean time as the lapse switching from $N_p=1$ to $N_p = \mp i,$ implying that the total time $T \equiv \int N_p dt$ becomes complex valued. Other, yet equivalent, representations of the geometry can then be obtained by choosing different ``paths'' to reach $T.$ This freedom has been exploited to relate the no-boundary proposal to the AdS/CFT correspondence \cite{Hertog:2011ky}, to find highly anisotropic instantons \cite{Bramberger:2017rbv} and to construct ekpyrotic no-boundary instantons \cite{Battarra:2014xoa, Battarra:2014kga}. 

For our discussion below it will be useful to rewrite the HH and V geometries for our choice of metric variables \eqref{eq:ansatz}. This can be achieved via the two choices
\begin{align}
\sinh(Ht_p) = H^2 N_{HH} t + i\,,\qquad \sinh(Ht_p) = H^2 N_{V} t - i\,,
\end{align}
where $N_{HH,V}$ will turn out to be the saddle point values of the lapse integral corresponding to the Hartle-Hawking and Vilenkin geometries respectively. These are explicitly given by
\begin{align}
N_{HH} =  \frac{\sqrt{H^2 q_1 - 1}}{H^2} - \frac{i}{H^2} \,,\qquad N_V =  \frac{\sqrt{H^2 q_1 - 1}}{H^2} + \frac{i}{H^2}\,, \label{HHsaddles}
\end{align}
where the final scale factor value is fixed to be $q(t=1)=q_1.$ We could also have considered the time reverses of these geometries, for which the real parts of $N$ would have had the opposite sign, in the same way as one could have let the physical time coordinate $t_p$ run over negative values. In our variables, the HH and V metrics may be written as
\begin{align}
\bar{q}(t) = H^2 N_{HH,V}^2 t^2 + (q_1-H^2 N_{HH,V}^2) t \,, 
\end{align}
with $0 \leq t \leq 1.$ Note that we have made use of the freedom of ``path'' to reach the final time $T$ with a constant (necessarily complex) value of the lapse $N.$ The imaginary parts of $N_{HH,V}$ again encode the direction of Wick rotation
\begin{align}
\frac{\dot{\bar{q}}}{2N_{HH}}\mid_{t=0} = +i\,,\qquad \frac{\dot{\bar{q}}}{2N_{V}}\mid_{t=0} = -i\,,
\end{align}
and hence they determine whether small fluctuations around these geometries are stable or not.

Motivated by the instability associated with the ``wrong'' Wick rotation, we may try to sum only over geometries that contain the ``correct'' Wick rotation. We can do this by fixing the initial expansion rate both to be Euclidean and to possess the appropriate sign. Fixing the expansion rate requires a Neumann condition, which we have seen to arise automatically from the Einstein-Hilbert action, without having to introduce a boundary term. We will see below that, perhaps surprisingly, adding no boundary term will not lead to Neumann conditions for all metric deformations, but it does so for the scale factor. In fact, we will see that adding no boundary term at all leads to a viable implementation of the no-boundary idea. The idea for such a ``no boundary term'' proposal was already mentioned by Louko and Halliwell in early papers on the subject \cite{Louko:1988bk, Halliwell:1988ik, Halliwell:1990tu} and recently in \cite{DiazDorronsoro:2018wro}, but it has not been analysed in much detail so far. 

At first, for simplicity, we will take the final condition to be given by a Dirichlet condition, fixing the size of the universe on the final hypersurface to $q(t=1)=q_1.$ Hence our boundary conditions are
\begin{align}
\frac{\dot{q}}{2N} \mid_{t=0} = + i\,, \qquad q \mid_{t=1} = q_1\,.
\end{align}
With these boundary conditions the equation of motion for $q$ is solved by
\begin{align}
q(t) &= H^2 N^2 t^2 + 2iNt + q_1 - 2iN - H^2 N^2\,, \label{NDq}
\end{align}
implying that after integrating over $q$ as described in the previous section, we are left with an action solely dependent on the lapse $N,$
\begin{equation}
\frac{S}{V_3} =  H^4 N^3 + 3 i H^2 N^2 - 3 H^2 q_1 N  - 3 i q_1\,. \label{NDaction}
\end{equation}
There are a few points to note about the form of this action: the first is that the action is explicitly complex. This is of course a direct result of imposing a Euclidean momentum at $t=0.$ This means however that even for real values of the lapse the weighting of the action (given by minus its imaginary part) will be non-zero. Picard-Lefschetz theory prescribes that a relevant saddle point of a path integral must be reached by flowing down from the original integration contour. For a purely real action this would imply that only saddle points with negative weighting could contribute, and this would preclude the Hartle-Hawking saddle point (which has a positive weighting $+4\pi^2/(\hbar H^2)$) from ever being relevant. The complexity of the action evades this obstruction. 

The second point to note about the action is that it does not contain a singularity at $N=0.$ Physically, this may be understood as follows: since we are fixing the initial momentum, and not the initial size of the geometries that are summed over, we are in effect summing over geometries of all possible initial sizes. This will include a geometry of size $q_1$ at $t=0,$ and the transition from this 3-geometry to the final hypersurface which also has $q=q_1$ can occur instantaneously, i.e. with $N=0.$ There is thus nothing singular occurring at $N=0$ \footnote{Note that, if the integration contour was taken to run from zero to infinity (to compute a propagator), the absence of a singularity would generate difficulties in the application of Picard-Lefschetz theory.}.

\begin{figure}
	\centering
	\includegraphics[width=0.6\textwidth]{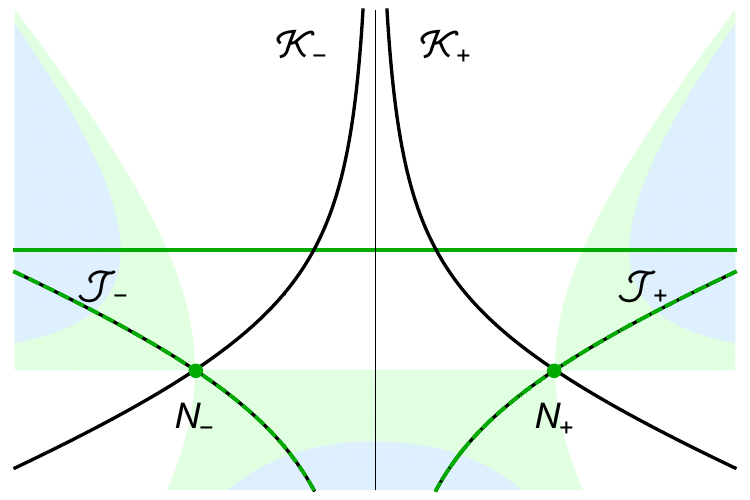} 
	\caption{Flow lines in the complex plane of the lapse function $N$ for the action \eqref{NDaction}, obtained by imposing boundary conditions with a Euclidean momentum at $t=0$ and a fixed final size at $t=1$ (the parameters used are $H=1$,  $\dot{q}_0/2N=i,$ $q_1=2$). The figure shows the paths of steepest ascent (black, ${\cal K}_\pm$) and steepest descent (black, ${\cal J}_\pm$) emanating from the saddle points (in green). Regions of asymptotic convergence are shown in blue, while regions of descent from the saddles are in light green. The dark green real line contour can be deformed into the dark green dashed thimbles ${\cal J}_- + {\cal J}_+.$}
	\label{fig:NeumannFlow}
\end{figure}

A third observation is that the action \eqref{NDaction} only contains two saddle points, located at
\begin{align}
N_\pm = \pm \frac{\sqrt{H^2 q_1 - 1}}{H^2} - \frac{i}{H^2}
\end{align}
These are precisely the Hartle-Hawking saddle points \eqref{HHsaddles}. Compared to the calculation starting from zero size \cite{FLT1}, which contained $4$ saddle points, the Neumann condition has eliminated the Vilenkin saddle points. Moreover, as Fig. \ref{fig:NeumannFlow} shows, the paths of steepest ascent/descent from these saddle points are such that they are both relevant to the path integral with Lorentzian contour of integration for the lapse. The real line contour for $N$ can in fact be deformed into the sum over both steepest descent paths ${\cal J}_+$,  ${\cal J}_-,$ while the arcs at infinity linking the thimbles to the real line yield zero additional contribution \cite{FLT1}. Note that we could not have used a Euclidean integration contour, as this would have been divergent at large positive imaginary values of the lapse. For the Lorentzian contour, the saddle point approximation then yields the result
\begin{align}
\Psi &\simeq e^{iS(N_-)/\hbar} + e^{iS(N_+)/\hbar} \nonumber \\
& = e^{+\frac{2V_3}{\hbar H^2}} \cos[  \frac{2 V_3 H}{\hbar} (q_1 - \frac{1}{H^2})^{3/2}]\,.\label{PsiHH}
\end{align}
Thus we have recovered the Hartle-Hawking wavefunction \eqref{PsiHH} from a Lorenztian path integral. 

Before discussing some implications of this result, it is interesting to see what happens if we try to extend the ``no boundary term'' prescription to more general metrics. An obvious class of importance are anisotropic metrics. For instance, we can consider the Bianchi IX metric
\begin{align}
ds_{IX}^2 = - \frac{N^2(t)}{q} dt^2 + \sum_m \left( \frac{l_m(t)}{2} \right)^2 \sigma_m^2\,,
\end{align}
with the $\sigma_m$ being one-forms on the 3-sphere. The $l_m$ are direction-dependent scale factors. It is useful to rewrite them as 
\begin{align}
l_1(t) =\sqrt{q} e^{\frac{1}{2}\left(\beta_+(t) + \sqrt{3}\beta_-(t)\right)}\,, \quad l_2(t) = \sqrt{q} e^{\frac{1}{2}\left(\beta_+(t) - \sqrt{3}\beta_-(t)\right)}\,, \quad
l_3(t) = \sqrt{q} e^{-\beta_+(t)}\,,
\end{align}
such that $q$ denotes the average scale factor squared, while the $\beta_\pm$ functions parameterise the deformations/squashings of the spatial slices (the conventional subscript $ \pm $, indicating the polarisation of the gravitational wave, should not be confused with the label of the two relevant saddle points). In these variables the action is given by
\begin{align}
S = V_3 \int dt N \left[  \frac{3}{4N^2}\left( 2 q \ddot{q} + \dot{q}^2  + q^2(\dot{\beta}^2_+ + \dot{\beta}^2_-) \right)  - \left( q  \Lambda + U(\beta_+, \beta_-)\right)\right]\,, \label{actionB9}
\end{align}
with the anisotropy potential
\begin{align} \label{anisotropypotential}
U(\beta_+, \beta_-)  & = - 2 \left( e^{ 2 \beta_+ } + e^{-\beta_+ - \sqrt{3}\beta_-} + e^{-\beta_+ + \sqrt{3}\beta_-} \right) + \left( e^{ -4 \beta_+ } + e^{2\beta_+ - 2\sqrt{3}\beta_-} + e^{2\beta_+ + 2\sqrt{3}\beta_-} \right) \nonumber \\ & = -3 + 6\left(\beta_-^2 + \beta_+^2 \right) + {\cal{O}}(\beta_\pm^3)\,.
\end{align}
Inspection of the action \eqref{actionB9} shows that something interesting has happened: in contrast to $q,$ the anisotropy parameters $\beta_\pm$ only appear with at most single derivatives in the action. Variation of the action thus automatically leads to a Dirichlet condition $\delta \beta_\pm=0$ on the boundaries, without the need to add a Gibbons-Hawking-York term. In short, the ``no boundary term'' prescription leads to a Neumann condition for the scale factor, but to Dirichlet conditions for the anisotropies.  

The equations of motion and constraint following from varying the action are
\begin{align}
& \ddot{q} + q\left( \dot{\beta}^2_+ + \dot{\beta}^2_- \right)  -\frac{ 2N^2}{3}\Lambda  = 0\,, \label{eqq} \\
& \ddot{\beta}_\pm + 2\frac{\dot{q}}{q}\dot{\beta}_\pm  + \frac{2N^2}{3 q^2} U_{,\beta_\pm} = 0\,, \label{eqbeta} \\
& \frac{3}{4}\dot{q}^2 = \frac{3}{4} q^2 \left( \dot{\beta}^2_+ + \dot{\beta}^2_-  \right)  +  q N^2 \Lambda + N^2U(\beta_+, \beta_-)\,. \label{constraint}
\end{align}
By multiplying \eqref{eqbeta} through by $q^2$ and solving near $q=0$ one can immediately see that a regular solution requires $U_{,\beta_\pm}=0$ at $q=0,$ which, if we assume that the saddle point will still be the Hartle-Hawking one with $q(t=0)=0,$ translates into the requirement $\beta_\pm(0)=0.$ This is indeed a Dirichlet condition, which moreover ensures that the constraint \eqref{constraint} can also be satisfied at the South Pole. 

At linear order in the anisotropies, the equation of motion \eqref{eqbeta} reduces to the equation for a linear gravitational wave,
\begin{equation}
\ddot{\beta}_\pm + 2 \frac{\dot{q}}{q} \dot{\beta}_\pm + \frac{8N^2 }{q^2} \beta_\pm = 0\,.  \label{eqpert}
\end{equation}
Let us now focus, for simplicity, on the background geometry associated with the saddle point $ N_+ $.
With the Dirichlet conditions $\beta_\pm(0)=0,\,\beta_\pm(1)=\beta_{1\pm}$ and neglecting backreaction, the solution is given by
 $\beta_\pm(t)=\beta_{1\pm} \cdot g(t)/g(1)$ with
 \begin{equation}
 g(t)= \left(t^2 (\sqrt{H^2 q_1 -1} - i )+4it\right)\left(t (\sqrt{H^2 q_1 - 1}- i )+2i\right)^{-2}\,.
 \end{equation}
The quadratic action associated with this solution (at large final $q_1$) is 
\begin{align}
S^{(2)}_\pm &= \frac{V_3}{2} \int_0^1 dt \, N \Bigl[q^2 \frac{\dot{\beta}_\pm^2}{N^2} - 8 \beta^2_\pm \Bigr] \\
 & = -\beta_{1\pm}^2  \frac{ 4 V_3 q_1}{3i + \sqrt{H^2 q_1 - 1}} \nonumber \\ &  = i \, \beta_{1\pm}^2 \frac{12 V_3}{H^2} - \beta_{1\pm}^2 \frac{4 V_3 \sqrt{q_1}}{H} + \mathcal{O} \Bigl( \frac{1}{\sqrt{q_1}} \Bigr)
\end{align} 
thus explicitly verifying that the Hartle-Hawking saddle points lead to stable, Gaussian distributed, perturbations in a Bianchi IX spacetime.


There remain two issues that require some discussion though: the first is one of interpretation. Based on the no-boundary geometry depicted in Fig. \ref{fig:Wick}, Hartle and Hawking proposed that the no-boundary wavefunction could be defined as a path integral where the sum over geometries is restricted to be over compact and regular metrics \cite{Hartle:1983ai}. It is very difficult to implement a sum over regular geometries only \cite{Halliwell:1989vu, DiTucci:2018fdg} -- we will return to this point below. For now, let us note that a direct implementation of the sum over compact geometries can be achieved by using the Dirichlet condition that the scale factor vanishes at $t=0,$ i.e. $q(t=0)=0.$ This has been shown to ultimately fail \cite{FLT2, Feldbrugge:2017mbc} in the sense that it leads to unstable perturbations. Here we have overcome this problem by fixing the initial momentum, rather than the initial size. Due to the uncertainty relation between scale factor and momentum this implies that the path integral sums over geometries with \emph{all} possible initial sizes, with some being larger than the current universe. The dominant geometry, of course, remains the HH geometry starting from ``nothing'', but off-shell all initial sizes are included. Hence, as defined here, the interpretation of the no-boundary wavefunction must change. In the present definition, imposing an appropriate Euclidean momentum takes precedence over the criterion of compactness. In the same spirit, it becomes questionable whether the no-boundary wavefunction truly describes tunneling out of nothing. Rather, as already discussed in \cite{DiTucci:2019dji}, it may describe a quantum transition from a prior state. If this is the case, then it may not be an ultimate theory of initial conditions. Despite this potential drawback, the no-boundary wavefunction retains highly appealing and non-trivial physical properties: in particular, it can describe how the universe becomes classical and provide an explanation for the Bunch-Davies state.

\begin{figure}
	\centering
	\includegraphics[width=0.6\textwidth]{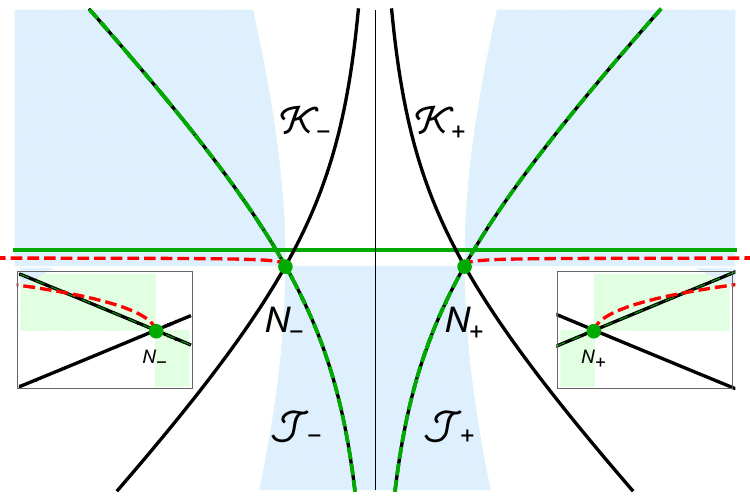}  
	\caption{The curves of zeroes (red, dashed), indicating the loci of geometries in which the scale factor vanishes at some time, emanate almost vertically from the saddle points (see inset) and then become horizontal, necessarily crossing the thimbles. The parameters used in making this figure are $H=1, \dot{q}_0 /2 N=i, q_1=100$.
	}
	\label{fig:Neumann}
\end{figure}

The second issue to be discussed is that of the potential singularities in the off-shell geometries that are being summed over. It is not entirely clear whether one should avoid off-shell singularities. On the one hand, they should be present since one is summing over all geometries, and already in quantum mechanics most paths that one sums over are ``singular'' as they are not differentiable. Moreover, if the singular geometries lead to infinite action they may automatically have zero weighting and thus provide no contribution to the path integral. On the other hand one may desire the integrations along Lefschetz thimbles to be mathematically well-defined and singularity free. Physically, the presence of singularities leaves the geometries in question subject to large corrections when higher-order terms are included in the action, thus rendering the calculation potentially unreliable. A singularity will occur when the scale factor $q(t)$ vanishes somewhere along the geometry, i.e. for some real $t$ with $0 < t \leq 1.$ At the saddle point $N_+$ itself  (and similarly at $N_-$) the geometry starts out at $q(0)\mid_{N_{+}}=0$ and it is regular there by construction. But nearby we may expect the singularity to occur at a small value $\delta t$ (note that off-shell in $N$ the constraint, which ensures regularity, is not satisfied). Starting from \eqref{NDq}, a short calculation shows that
\begin{align}
q(\delta t)\mid_{N_+ + \delta N} = 0 \quad  \rightarrow \quad \delta N =  \frac{iN_+}{\sqrt{H^2 q_1 -1}} \delta t\,.
\end{align}
Thus the ``curve of zeroes'', representing the locus of geometries containing a singularity, emanates from the saddle point at an angle which is a rotation by $\pi/2$ compared to the angle subtended by the saddle point itself. At large $q_1,$ the saddle point is almost real and hence the curve of zeroes leaves the saddle point approximately vertically (in the positive imaginary lapse direction). This is confirmed by the numerical calculation shown in Fig. \ref{fig:Neumann}.  At large $N$ (for singularities occurring near $t=1$) the curve of zeroes runs off almost horizontally, just below the real $N$ line and with approximate imaginary part given by $-i/(2H^2).$

Meanwhile the thimbles are defined via the relation $Re(S(N))=Re(S(N_s)),$ since they correspond to stationary phase paths of the integrand. Again perturbing away from the saddle point, one finds that for small deviations $\delta N$ the thimble obeys the relation 
\begin{align}
\text{Re}\left( \frac{1}{2}S_{,NN}(N_+)(\delta N)^2\right)=\text{Re}\left( 3H^2 \sqrt{H^2q_1 -1}(\delta N)^2\right)=0\,.
\end{align}
This equation is quadratic in $\delta N$ because the first derivative of the action vanishes at the saddle point. Since $q_1 > 1/H^2$ we must have $(\delta N)^2$ purely imaginary, or in other words we need $\delta N$ to point in the directions $e^{i\pi/4},e^{i3\pi/4}, e^{i5\pi/4}, e^{i7\pi/4}.$ By inspection we can see that the thimble points at an angle $\pi/4$ away from the saddle point. Thus the thimble necessarily intersects the curve of zeroes, and will necessarily contain at least one singular geometry. Moreover, the deformation of the original integration contour along the real $N$ line will have to pass through the curve of zeroes to reach the thimble. As we said above, it is not entirely clear whether this is a serious obstruction or not. This potential obstruction can in any case be resolved in interesting ways, as we will demonstrate in the following.

\section{The final Hubble rate as a boundary condition}\label{sec:hubblerate}

We have seen in the previous section how, by carefully fixing the initial momentum, it is possible to define a path integral peaked around the HH saddle point(s). However, we also saw that the thimbles that are integrated over contain singular geometries whose associated Ricci scalar diverges. As a result, this formulation is potentially sensitive to higher order quantum corrections, as terms involving higher powers of the Riemann curvature could significantly affect the integral. In this section we will provide a formulation of the quantum cosmology integral which avoids this problem making use of an initial Neumann condition and a final covariant Robin condition. The price to pay will be a redefinition of the integral over the lapse $ N $.  

It was shown in \cite{Krishnan:2017bte} that the Robin problem for gravity in 4 dimensions is obtained from  the boundary term
\begin{equation}
S_B = \frac{1}{\xi} \int_{\partial M} d^3 y \, \sqrt{h} \, ,\label{actionRob}
\end{equation}
where $\xi $ is a constant.  Here we are only going to consider a final Robin boundary condition, i.e. the Robin boundary term is evaluated on the final spatial surface. With the ansatz~\eqref{eq:ansatz} the Robin boundary term becomes $S_B = \frac{V_3}{\xi} q_1^{3/2}$ and, upon variation, the total action $ S=S_{EH}+S_B $ leads to the boundary conditions 
\begin{align}
\frac{\dot{q}_0}{2 N} &= \pi_0\,, \\
\frac{\dot{q}_1}{N } + \frac{2 \sqrt{q}_1}{\xi} & = \alpha  \label{rob} \, ,
\end{align}
with generic $\alpha$ and $\pi_0$. The solution to the equation of motion satisfying these boundary conditions is 
\begin{equation}
\overline{q}(t) = H^2 N^2 t^2 + 2 N \pi_0 t + \frac{\xi^2}{4} (\alpha - 2 (H^2 N + \pi_0))^2 - N (H^2 N + 2 \pi_0) \, ,
\end{equation} 
and the total classical action is 
\begin{align}
\frac{S}{V_3} = &N^3 H^4 (1 - H^2 \xi^2) + N^2 (\frac{3}{2} H^4 \xi^2 (\alpha - 2 \pi_0) + 3 H^2 \pi_0)  \nonumber \\
& + N (- \frac{3}{4} H^2 \xi^2(\alpha - 2 \pi_0 )^2 + 3  \pi_0^2  + 3) + \frac{\xi^2}{8} (\alpha - 2 \pi_0)^3  \, .
\end{align}
In our coordinates \eqref{eq:ansatz}, the Hubble rate is given by $\frac{\dot{q}}{2 N \sqrt{q}},$ and thus we can see from \eqref{rob} that if we set $\alpha=0,$ we may interpret $H_1=-\frac{1}{\xi}$ as the Hubble rate on the final hypersurface. Note that due to the closed spatial slicing of de Sitter space we should require that $H_1 \le H$ or, equivalently, $\xi^2 H^2\ge 1$. With vanishing $\alpha,$ the action can be usefully rewritten as
\begin{equation}
\frac{S}{V_3} = H^2 (1 - H^2 \xi^2) (N + \frac{\pi_0}{H^2})^3 + 3 (N + \frac{\pi_0}{H^2})  - \pi_0 \frac{3  + \pi_0^2}{H^2}\,.\label{actionoffset}
\end{equation}
It is clear from this expression that for real boundary conditions $\pi_0 \in \mathbb{R}$, the integrand $e^{i S}$ oscillates along the real $N$ line and is conditionally convergent. Note that, since $(1 - H^2 \xi^2) \le 0 $, the asymptotic regions of convergence for the lapse integral lie in the wedges between the angles $(\frac{\pi}{3},\frac{2\pi}{3})$, $(\pi,\frac{4\pi}{3})$ and $(\frac{5\pi}{3},2\pi)$. Thus the Lorentzian integral can be defined and calculated, using Picard-Lefschetz theory to deform the real $N$ line to the appropriate steepest descent contour \cite{FLT1}. For example, with the boundary condition $\pi_0=0$ the saddle points are located on the real axis at $N_{\pm} = \pm \frac{H_1}{H^2 \sqrt{H^2-H_1^2}}$ and they describe the expansion of the universe from the waist of the de Sitter hyperboloid where $q(t=0)=1/H^2$ to a final hypersurface with Hubble rate $H_1=-1/\xi,$ according to $q(t)=H^2 N_\pm^2 (t^2-1+\frac{H^2}{H_1^2}).$

\begin{figure}
	\centering
	\includegraphics[width=0.45\textwidth]{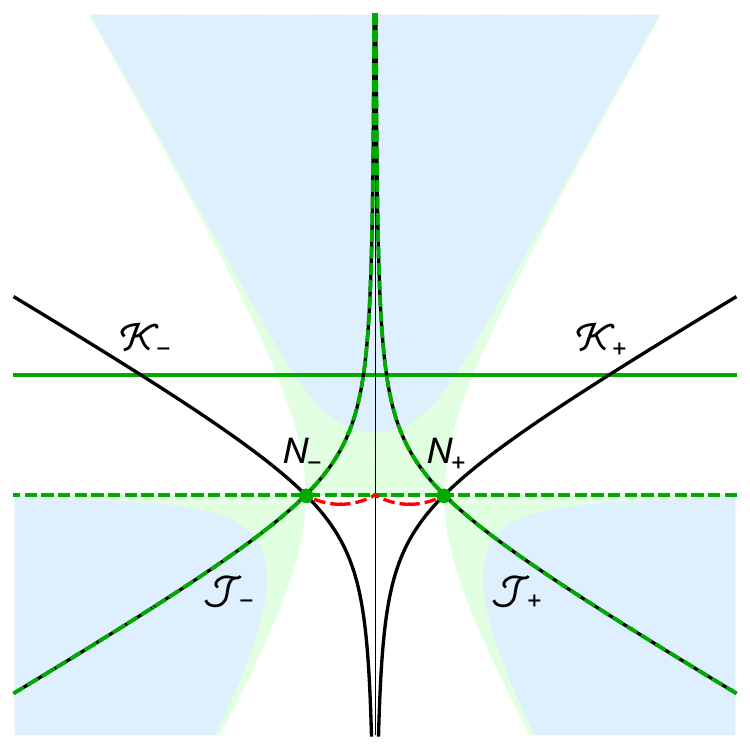} 
\includegraphics[width=0.45\textwidth]{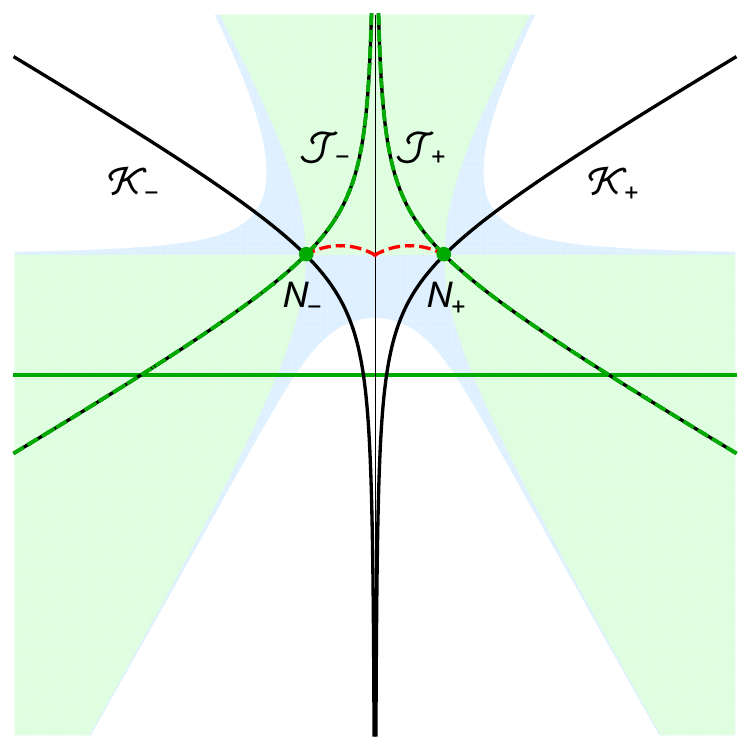}
	\caption{Flow lines and saddle points in the complex lapse plane, for a Neumann condition on $\Sigma_0$ and a covariant Robin condition on $\Sigma_1.$ {\it Left panel:} no-boundary wavefunction, $ \pi_0=i $.
	{\it Right panel:} Tunneling wavefunction, $ \pi_0=-i $. In both cases we set $H=1, H_1=1/2.$ In solid green, the real $N$ line; in dashed green, the shifted defining contour. We show in light green the regions of descent from the saddles and in light blue the regions where $ \text{Re}(iS)<0 $. The deformed contour runs along the steepest descent lines $ J_\pm $ and is shown in dashed green superimposed on the black steepest descent contours. The lines of zeroes (in red, dashed) are always avoided by the thimbles.
	}
	\label{fig:RobinFlow}
\end{figure}

For the no-boundary wavefunction however we need an imaginary initial momentum. When $\pi_0 \in i \mathbb{R}$ the integrand is oscillatory not along the real $N$ line, but rather (asymptotically) along the line parallel to the real axis with an offset given by $-\frac{\pi_0}{H^2},$ as can be seen by inspection from the action \eqref{actionoffset}. In fact, if $\pi_0 \in i \mathbb{R}^-$, the integral along the real line is convergent whereas for $\pi_0 \in i \mathbb{R}^+$ it explicitly diverges for large real $N$. Consequently, if we were to impose the Vilenkin momentum $\pi_0 = - i$ or any classically allowed boundary condition, the Lorentzian path integral would be mathematically well defined -- see also Fig. \ref{fig:RobinFlow}. However in order to implement the no-boundary wavefunction we need $\pi_0 = + i $, for which the integral along the real line diverges. 

A meaningful integral can be obtained by shifting the defining contour to the line $N=-\frac{i}{H^2} + x$ with $x \in \mathbb{R}$ (or potentially shifting the defining contour even below this line). This represents a departure from the exact Lorentzian integral, which is forced upon us by requiring the integral to be well defined. In some sense the departure is quite minimal, as the integration direction is still in the Lorentzian time direction. However, it is a clear departure as the defining sum is now over complex geometries. The extent to which this might constitute a problem may be debated. If we assume this new contour of integration, then the path integral will be equivalent to a sum over the two associated Lefschetz thimbles, as shown in Fig. \ref{fig:RobinFlow}. The thimbles are peaked on the HH saddle points
\begin{align}
N_{\pm} = \frac{1}{H^2}\left(\pm \frac{H_1}{\sqrt{H^2-H_1^2}} -i\right)\,,
\end{align} 
and consequently we obtain the approximate wave function
\begin{align}
\Psi \simeq e^{iS(N_-)/\hbar} + e^{iS(N_+)/\hbar}  = e^{+\frac{2V_3}{\hbar H^2}} \cos\left[\frac{2V_3}{\hbar H^2}(q_1 H^2 - 1)^{1/2} \right]\,,
\end{align}
where we note that the phase of the wave function only grows linearly with the scale factor\footnote{We thank Vikramaditya Mondal for pointing out an error in an earlier version of this paper.}. As the contributions from the two saddle points are both of WKB form, we still expect decoherence to occur, though presumably at a slower rate than for the standard Hartle-Hawking wave function \eqref{PsiHH}, in which the phase grows with the spatial volume. This point deserves further study in the future.

In analogy with the discussion of the previous section, we would now like to know whether the thimbles intersect any geometry that contains a singularity in the form of $q(t)=0$ for some real $t$ with $0<t<1.$  The equation $q(t) =0 $ is solved for 
\begin{equation}
t_{1,2} = - i \frac{H_1 \pm i \sqrt{H^2 - H^2_1} (1 - i H^2 N)}{H^2 N H_1}
\end{equation}
The imaginary part of $t_{1,2}$ vanishes respectively on the two circles $m^2 + n^2 + \frac{m}{H^2} \pm \frac{n H_1}{H^2 \sqrt{H^2 - H_1^2}} = 0$, where $ N=n + i m $. 
Moreover, the real part of $t_{1,2}$ should vary between $0$ and $1$, which imposes the condition $0< -\frac{H_1 m+\sqrt{H^2-H_1^2} \, n}{H^2 H_1 \left(m^2+n^2\right)} <1 $. This condition selects the arcs of the circumferences that link the saddle points (where $t _{1,2}= 0 $) to the point $ (n,m) = (0 , - \frac{1}{H^2})$ (where $t_{1,2} = 1$). 
Combining  the two conditions above, it is easy to see that the line of zeroes corresponds to the lower arcs of the two circles emanating from the saddles points, see Fig.~\ref{fig:RobinFlow}. In particular, one can verify that close to the saddle points the conditions impose $ m\leq \text{Im}(N_{\pm})=-1/H^2 $. Near the saddle point the curve of zeroes is approximated by the straight line
\begin{align}
q(\delta t)\mid_{N_{\pm} + \delta N} = 0 \rightarrow \delta N = \delta t  \, \frac{H_1^2}{(H^2 - H_1^2)} ( - \frac{i}{H^2} \mp \frac{\sqrt{H^2 - H_1^2}}{H^2 H_1})\,.
\end{align}
Thus at the two saddle points the line of zeroes forms an angle of $\tan(\theta) =\pm \frac{H_1}{\sqrt{H^2 - H_1^2}}$ with the horizontal, respectively. In other worlds, $ \theta \in (\pi, 3 \pi /2) $ for $ N_+ $ and $ \theta \in (0, -\pi/2) $ for $ N_- $.

As for the thimble, it is given by the equation
\begin{align}
\text{Re}\left( \mp 3\frac{H^2}{H_1^2} \sqrt{H^2-H_1^2}(\delta N)^2\right)=0 \, .
\end{align}
Since $H_1\le H$, the flow lines again point in the directions $e^{i\pi/4},e^{i3\pi/4}, e^{i5\pi/4}, e^{i7\pi/4}.$ In this case however the steepest descent path points at $3\pi/4$ radians away from the saddle point $N_+$ (and $\pi/4$ from $N_-$). 
Thus the thimble and line of zeroes avoid each other, and numerical calculations confirm this beyond leading order for the entire trajectory traced by these lines (see Fig.~\ref{fig:RobinFlow}).

In summary, the no-boundary wavefunction can be defined with an initial Neumann condition and a final covariant Robin condition. The latter has the physical interpretation of fixing the  Hubble rate on the final slice. The thimbles associated with the HH saddles also avoid singular geometries everywhere, so that the curvature is everywhere bounded and general relativity can be trusted at every step. The price to pay is a redefinition of the lapse integral. With the standard choice, a contour coinciding with the real line, the integral would have been divergent. The integral is convergent if the contour is shifted by an imaginary offset of at least $-i/H^2$. The important feature compared to the definitions that were explored in earlier works is that the initial Neumann condition eliminates the saddle points with unstable fluctuations, leaving only the stable HH saddles. Meanwhile, the physically attractive final condition of imposing the current Hubble rate rather than the (currently unobservable) size of the universe eliminates any potential interference of singular geometries.


\section{Canonical Robin boundary conditions} \label{sec:canonical}

In the previous section the Robin boundary condition was implemented covariantly. Once the defining contour was shifted below the real axis, we were then able to recover the no-boundary wavefunction and moreover avoid any potential ambiguities regarding sums over singular geometries.  In this section we will discuss a different way of imposing boundary conditions that relate field values and momenta. These ``canonical'' Robin conditions will preserve the convergence of the path integral when summed over Lorentzian geometries. 

We study a path integral with Robin boundary conditions at both boundaries, $t=0 $ and $t=1$, by considering the canonical Robin boundary term,
\begin{equation}
\frac{S_B}{3 V_3}= + \alpha_0 q_0 + \frac{(q_0 - q_i)^2}{2 \beta_0} + \alpha_1
 q_1 + \frac{(q_1 - q_f)^2}{2 \beta_1} \, ,
\end{equation}
with arbitrary $\alpha_0$, $ \alpha_1$, $ \beta_0 $, $\beta_1$, $ q_i $, $ q_f $. This is a generalization of the canonical boundary conditions discussed in \cite{DiTucci:2019dji}\footnote{Compared to \cite{DiTucci:2019dji} the $\alpha$s and $\beta$s are rescaled by a factor $3V_3$ here.}. Analogous boundary conditions were studied in \cite{Vilenkin:2018dch} in the context of the tunneling proposal, and in \cite{DiTucci:2019xcr} and  \cite{Bramberger:2019zks} with applications to the problem of the Bunch-Davies vacuum in inflationary spacetime, and to the problem of inflaton jumps. 

This type of Robin condition is different from what we considered so far and in particular it is not covariant.  
Despite this drawback in what concerns diffeomorphisms, canonical Robin conditions have nice properties from a quantum mechanical point of view. The boundary terms can in fact be interpreted as initial and final states of (in general complexified) Gaussian form.  In the 
Dirichlet ($\beta_{0,1} \rightarrow 0 $) and Neumann limit ($\beta_{0,1}  \rightarrow \infty $) the initial/final state reduces to a delta function and a plane wave respectively. In other words, it reproduces what in the canonical quantisation framework would be eigenstates of the `position' operator $\hat{q}$ and `momentum' operator $\frac{\hat{\dot{q}}}{2 N}$ with eigenvalues $q_i $ ($q_f$) and $ \alpha_ 0  $ ($a_1$). 

The bulk action is the action for the Dirichlet problem for gravity (including the GHY boundary term),
\begin{equation}
S_{EH}+ S_{GHY}= 3 V_3 \int_0^1 dt \, [ - \frac{\dot{q}^2}{4 N} + N (1 - H^2 q)] \, .
\end{equation} 
The variation of the total action with respect to $q$ reads
\begin{equation}
\frac{\delta S }{3 V_3}=  \int_0^1 dt [\frac{ \ddot{q}}{2 N} - H^2 N ] \delta q + [- \frac{\dot{q_1}}{2 N }  + \alpha_1 + \frac{(q_1 - q_f)}{\beta_1}] \delta q_1 + [\frac{\dot{q}_0}{2 N}  + \alpha_0 + \frac{(q_0 - q_i)}{\beta_0}] \delta q_0 \, ,
\end{equation}
and implies the familiar equation of motion~\eqref{eomq} supplemented with the boundary conditions
\begin{equation}\label{eq:canrobcanrobboundary}
 \frac{\dot{q}_0}{2 N}  + \alpha_0 + \frac{(q_0 - q_i)}{\beta_0} = 0 , \quad  - \frac{ \dot{q_1} }{2 N }+ \alpha_1 + \frac{(q_1 - q_f)}{\beta_1} = 0 \, .
\end{equation}
From these we can clearly see that generic Dirichlet boundary conditions are recovered in the limit $ \beta_{0,1} \rightarrow 0 $; Neumann conditions are recovered for $ \beta_{0,1} \rightarrow \infty $. The unique solution to the equation of motion satisfying these boundary conditions is 
\begin{align}\label{eq:cancanqt}
\bar{q}(t) = &H^2 N^2 t^2 + \left( \frac{2 N (\alpha_1 \beta_1 - \alpha_0 \beta_0   + H^2 N^2 - \beta_1 H^2 N - q_f + q_i)}{ (\beta_1 + \beta_0) - 2 N } \right) t  \\
&+  \frac{1}{(\beta_1 + \beta_0) - 2 N } \left(\beta_0 \left(2 N \alpha_0 -  \beta_1 (\alpha_1 + \alpha_0) - H^2 N^2 +  \beta_1 H^2 N + q_f\right) + (\beta_1 - 2 N) q_i\right)  \, . \nonumber
\end{align}
Performing the path integral over the scale factor by shifting variables to $q=\bar{q} +Q$, up to an unimportant prefactor the path integral reduces to an ordinary integral over the lapse function $\Psi \simeq \int dN e^{iS/\hbar}$ with effective action
\begin{align}
S = & \frac{3 V_3}{12 N - 6 (\beta_1 + \beta_0)} \Bigg(   H^4 N^4 - 2 (\beta_1 + \beta_0) H^4 N^3 + \nonumber \\
&+ 3 N^2 \left( -2 H^2 \left(-\alpha _0 \beta _0-\alpha _1 \beta _1+q_f+q_i\right)+\beta _0 \beta _1 H^4+4 \right)  \nonumber\\
& + 6 N \left( q_f \left(2 \alpha _1+\beta _0 H^2\right)-\left(\alpha _0^2+1\right) \beta _0 -\beta _1 \left(\alpha _1^2+\left(\alpha _0+\alpha _1\right) \beta _0 H^2+1\right)+q_i \left(2 \alpha _0+\beta _1 H^2\right) \right) \nonumber\\
& -3 \left(q_f-q_i\right){}^2-6 \left(\alpha _0+\alpha _1\right) \beta _0 q_f+3 \left(\alpha _0+\alpha _1\right) \beta _1 \left(\left(\alpha _0+\alpha _1\right) \beta _0-2 q_i\right) \Bigg)  \, .
\end{align}
The singularity of the action, located at $N = 0 $ for Dirichlet boundary conditions and absent for Neumann boundary conditions, is now at $N_* = \frac{ (\beta_0 + \beta_1) }{2}$. To evaluate the remaining integral over $ N $, we will have to study the four saddle points of the action,
\begin{align}
N_{c_1 ,c_2 } =  &\frac{ (\beta_0 + \beta_1) }{2} + \frac{c_1}{H^2} \sqrt{- 1 - H^2 \alpha_0 \beta_0 + \frac{ H^4 \beta_0^2}{4} + H^2 q_i} \nonumber \\
&+ \frac{c_2}{H^2} \sqrt{-1 - H^2 \alpha_1 \beta_1 + \frac{ H^4  \beta_1^2 }{4}+ H^2 q_f} \, . \label{eq:canonicalall}
\end{align}
where $c_1, c_2 \in {-1 , 1}$. 

We will now focus our attention on the no-boundary wavefunction. For this, we require the  saddle point geometry to start out at zero size. The initial size at the saddle points is 
\begin{equation}
q_0=q(0) =q_i - \alpha_0 \beta_0 + \frac{\beta_0^2  H^2}{2}   + c_1  \beta_0 \sqrt{-1 - H^2 \alpha_0 \beta_0 + \frac{ H^4  \beta_0^2 }{4} + H^2 q_i} \, ,
\end{equation} 
As mentioned above, the condition $\beta_0 = 0$ corresponds to the Dirichlet limit for the initial boundary. Then, if $q_i = 0 $, the initial size vanishes for every value of $N$, not only at the saddle point. Away from the Dirichlet limit, the condition for the vanishing of the initial size is $ \alpha_0 = \pm i + q_i /\beta_0 $. We will still require $ q_i =0$, in the attempt of describing a universe that started from nothing. The two signs of $\alpha_0$ then implement the tunneling and the no-boundary geometry. For the no-boundary case we must choose $\alpha_0 = - i $. The saddle points then become
\begin{align}
N_{-1 , c_2} &= - \frac{i}{H^2} + \frac{ \beta_1 }{2 } + \frac{c_2}{H^2} \sqrt{-1 - H^2 \alpha_1 \beta_1 + \frac{  H^4  \beta_1^2}{4}  + H^2 q_f}  \, ,\\
N_{+1 , c_2} &= + \frac{ (2 \beta_0 + \beta_1)}{2} + \frac{i}{H^2} + \frac{c_2}{H^2} \sqrt{-1 - H^2 \alpha_1 \beta_1 + \frac{ H^4  \beta_1^2 }{4} + H^2 q_f}  \, .
\end{align}
These expressions fix our sign convention for the square root appearing in~\eqref{eq:canonicalall}. The saddles $N_{-1 , c_2}$  have vanishing initial size and therefore play the role of the HH saddles,
\begin{align}
q(0) \Bigr|_{N_{-1 , c_2}} &= 0 \, ,\\ 
q(0) \Bigr|_{N_{+1 , c_2}} &= \beta_0 (2 i +  H^2 \beta_0 ) \, .
\end{align}
Had we chosen $\alpha_0 = +  i $ instead, the initial size would have vanished at the Vilenkin saddle points $N_{+1 , c_2}$.

At this point, the parameter space spanned by the boundary conditions is still very large. To further constrain the parameters, we require that the boundary conditions describe complexified de Sitter geometries, $q(t) = H^2 N^2 t^2 + (q_1 - H^2 N^2)t$, $N = \pm \frac{\sqrt{H^2 q_1 - 1}}{H^2} - \frac{i}{H^2}$. This implies $\frac{\dot{q}_1}{2 N }  = \sqrt{H^2 q_1 - 1}$ and $\frac{\dot{q}_0}{2 N } =  + i $, which when plugged into~\eqref{eq:canrobcanrobboundary} impose
\begin{align}
- \sqrt{H^2 q_f - 1} + \alpha_1 + \frac{(q_1 - q_f) }{\beta_1} = 0  \,, \qquad
\alpha_0 = -i  \label{R1} \, .
\end{align}
In other words, it is not sufficient to impose $\alpha_0 = -  i$ but we must also require a relation between $\alpha_1$ and $\beta_1$ as in \eqref{R1}. Under this condition, we recover one of the HH saddles independently of $\beta_0$ and $\beta_1$,
\begin{equation} 
\begin{split}\label{eq:cansaddles}
N_{-1 , +1 } &= N_{HH} =   +\frac{\sqrt{H^2 q_f - 1}}{H^2} - \frac{i}{H^2} \, , \\ 
N_{-1 , - 1 } &=  - \frac{\sqrt{H^2 q_f - 1}}{H^2} - \frac{i}{H^2} + \beta_1 \, ,\\ 
N_{+1 , + 1 } &=   + \frac{\sqrt{H^2 q_f - 1}}{H^2} + \frac{i}{H^2}  +  \beta_0 \, ,\\
N_{+ 1 , -1 } &=  - \frac{\sqrt{H^2 q_f - 1}}{H^2} + (\beta_1 + \beta_0) + \frac{i}{H^2} \, .\\ 
\end{split} 
\end{equation}
Note that, when this condition is satisfied, the two geometries that start out at zero size have different final size and final velocity,
\begin{align}
q(1) \Bigr|_{N_{c_1 , +1}} &= q_f \, , \label{q11} \\
q(1) \Bigr|_{N_{c_1 , -1}} &= q_f +  \beta_1  (- 2\sqrt{H^2 q_f - 1} +  \beta_1  H^2 )\, , \label{q12}\\
\frac{ \dot{q}_1}{2 N} \Bigr|_{N_{c_1 , +1}} &= + \sqrt{H^2 q_f -1}  \, ,\label{eq:canRobvel1} \\
\frac{ \dot{q}_1}{2 N} \Bigr|_{N_{c_1 , -1}} &= - (\sqrt{H^2 q_f -1} -  \beta_1 H^2 ) \, .\label{eq:canRobvel2}
\end{align}
unless $\beta_1 =0 $ or $\beta_1 = \frac{2 \sqrt{H^2 q_f - 1}}{ H^2 }$. The 3-geometry on the final boundary is also real only if $\beta_1$ is real. As a consequence of the de Sitter condition \eqref{R1}, real $ \beta_1 $ implies real $\alpha_1$, as well. A real $
\alpha_1$ is desirable because in the Neumann limit $\alpha_1$ is the velocity on the final boundary.

\begin{figure}
	\centering
	\includegraphics[width=0.6\textwidth]{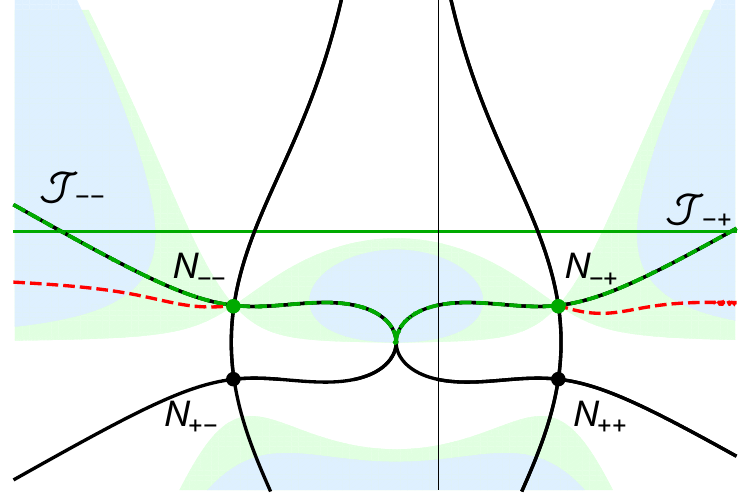}  
	\caption{The flow lines and saddle points in the canonical implementation of the Robin conditions. The parameters used in making the figure are $H=1$, $q_f=3$, $\beta_0= -3i$, $\beta_1=-1$, $\alpha_0$ and $\alpha_1$ according to Eq.~\eqref{R1}. Since here $ \beta_0 = i \bar{\beta} $, $ \bar{\beta}> 2/ H^2,$ the V saddle points have moved below the HH saddles, and a Stokes phenomenon has taken place, leaving only the two HH saddles (in green) as relevant. The original integration contour, i.e. the real line (green), can be deformed into the sum of the thimbles $ \mathcal{J}_{+-} $ and $ \mathcal{J}_{++} $ (green, dashed). Because $\beta_1 \neq 0,$ the symmetry across the imaginary lapse axis is shifted. The regions of convergence and descent from the relevant saddles are shaded in blue and green respectively. For these parameter values, the curves of zeroes (red, dashed) do not cross the thimbles, as explained in the main text. 
	}
	\label{fig:HHtorel}
\end{figure}

The analysis of the flow lines (steepest ascent/descent lines) proceeds in direct analogy to the earlier paper \cite{DiTucci:2019dji}, to which we refer for further details. We may start with the Dirichlet-Dirichlet limiting case $\beta_0 = \beta_1 = 0 $. This takes us back to the calculation discussed in great detail in \cite{FLT1}. In this case, one of the Vilenkin-like saddle points, $N_{+1 , +1 }$, is the relevant saddle if the integration contour is taken to be the positive real line, starting from the singularity at $ N_*=0 $. Both tunneling saddle points $N_{+1 , c_2}$ are relevant if the contour runs along the entire real line, avoiding the singularity from above. This remains true for all real $ \beta_0 $, $ \beta_1 $.
 
To make the HH geometry dominant, we need to consider imaginary or complex $ \beta $s. As explained below, we can interpret imaginary $ \beta_0 $ ($ \beta_1 $) as allowing for uncertainty in the initial (final) size. Demanding that the final size of the universe takes a real value requires $\beta_1$ to be real, as can be seen from \eqref{q12}, hence we will stick to this choice for now.

Turning on the uncertainty in the initial size $\beta_0$ while keeping $ \beta_1$ real, the upper saddle points move downwards if $\beta_0 $ is negative and imaginary, and the HH saddles become relevant for $\beta_0 = - i \overline{\beta} $ with $\overline{\beta} > 2/ H^2$, as was first shown (for $\beta_1=0$) in \cite{DiTucci:2019dji}, see also Fig.~\ref{fig:HHtorel}. This value for $\overline{\beta}$ is not surprising: the original position of the Vilenkin saddle points is at $\text{Im}(N)=+i/H^2,$ while the HH saddles reside at $\text{Im}(N)=-i/H^2.$ Thus $\beta_0$ must simply be large enough in magnitude to pull the V saddle points below the HH saddles, in effect moving them out of the way. As the V saddles move below the HH saddles, a Stokes phenomenon takes place and the V saddle points become irrelevant. Thus, for large enough $|\beta_0|,$ only the HH saddles remain relevant, implying that we have found another path integral formulation of the Hartle-Hawking wave function. For completeness, we show in Appendix \ref{WdWproof} that the path integral satisfies the Wheeler-DeWitt equation. 

The path integral is however again subject to the potential problem of the thimbles crossing singular geometries. As was shown in \cite{DiTucci:2019dji}, in order to avoid the ``curve of zeroes'', $|\beta_0|$ cannot take too large values.  For  example, for $\beta_1 = 0 $, the condition is that $|\beta_0| < 2 \Bigl( \frac{3 H^2 q_f -4}{H^2 q_f -2} \Bigr)$ as long as $q_f>2/H^2$. This bound can be found as follows.  The line of zeroes in the complex N plane is given by the two conditions $ \text{Re}(q(t))=0 $ and $ \text{Im}(q(t))=0 $ for $ 0\leq t\leq 1$. 
For the special parameters considered here, the vanishing of the imaginary part of Eq.~\eqref{eq:cancanqt} implies
\begin{equation}
t = \frac{\beta_0}{H^2 m } \frac{\beta_0 (1 + H^2 m) - i H^2 (m^2 + n^2) + i q_f}{4 n^2 + (i \beta_0 + 2 m )^2}\label{t} \, ,
\end{equation}
where $N = n + i m $ and we assumed that $\beta_0$ is imaginary. 
There is no concise analytic expression for the curve of zeros. However, all we need to know is the slope of the curve at the saddle point. It is therefore sufficient to plug Eq.~\eqref{t} into Eq.~\eqref{eq:cancanqt}, expand to first order around $N_{-1 , c_2}$ and then solve for the curve $m(n)$. 
The slope of this curve at the saddle points is found to be 
\begin{equation}
\tan(\phi) = c_2 \frac{(4 i + \beta_0 H^2) \sqrt{H^2 q_f - 1}}{8 i + H^2 (\beta_0 - 2 \, i \, q_f)} \, .
\end{equation} 

Defining $e^{i S} = e^{h + i s}$, the direction of the steepest descent path at the saddle point is given by the direction $\theta$ of the eigenvector of the Hessian matrix for the Morse function $h(n,m) = - \text{Im} (S (n , m))$, associated with the negative eigenvalue. For no-boundary values of the parameters, the action $S$ reads 
\begin{align}
S = \frac{V_3}{2 \left(2N-\beta _0\right)} \bigg(&H^4N^4 -2 \beta _0 H^4 N^3   +6 N^2 \left(2- H^2 \left(q_f+i \beta _0\right)\right)   \\ &+6 q_f N \left(\beta _0 H^2+2 \sqrt{H^2 q_f-1}\right) -6 \beta _0 q_f \left(\sqrt{H^2 q_f-1}-i\right)- 3 q_f^2 \bigg)\,. \nonumber
\end{align}
Then for the HH saddle points $N_{-1, c_2}$ one finds
\begin{equation}
\tan(\theta) =c_2 \frac{-2 + i \beta_0 H^2}{2\sqrt{H^2 q_f  - 1} + H \sqrt{4 q_f -4 i \beta_0 - H^2 \beta_0^2}} \, , \label{tetathimble}
\end{equation} 
where we assumed that $|\beta_0| > \frac{2}{H^2}$.  The thimble does not cross the line of zeroes a second time (besides the saddle point, where the zero scale factor point is regular by construction) if the condition $\tan(\theta) > \tan(\phi)$ is satisfied. 
For $q_f > \frac{2}{H^2}$ this happens for 
\begin{equation}
|\beta_0| < 2 \frac{ 3 H^2 q_f -4}{H^2 ( H^2 q_f -2)} \, .
\end{equation}
In the end, we are left with the following range for $\beta_0,$ where the HH saddles are the only relevant saddle points and where the thimbles are singularity free,
\begin{align}
\frac{2}{H^2} < |\beta_0| < 2 \frac{ 3 H^2 q_f -4}{H^2 ( H^2 q_f -2)} \approx \frac{6}{H^2}\,.
\end{align}

Let us now turn on $\beta_1$ once again, i.e. let us consider generic canonical Robin condition for the final boundary. If $ \beta_1 $ is real the two HH saddle points are both relevant for the contour $-\infty < N < \infty,$ as is clear from Fig. \ref{fig:HHtorel}. The quantum creation of the universe is therefore well approximated by the interference between two different geometries. The two relevant geometries have different final sizes,  \eqref{q11} and \eqref{q12}, and are not in general the time reversals of each other in the sense that $N \rightarrow - N $. However, for large $q_1$ the relative final size between the two saddles decreases,
\begin{equation}\label{key}
\frac{\Delta q(1)}{q(1)} \mid_{q_1\rightarrow\infty} \sim \frac{1}{\sqrt{q_1}} \, ,
\end{equation}
while the final velocities~\eqref{eq:canRobvel1},~\eqref{eq:canRobvel2} tend to opposite values, $ \pm \sqrt{q_1} $. 

The two HH saddle points have the same weighting $\text{Re}(i S) = \frac{2 V_3 }{ H^2}$, but different phases,
\begin{align*}
&\text{Im}\left(\frac{S _{-1 , +1}}{V_3}\right) =  2 \frac{  \sqrt{H^2 q_f - 1}}{H^2 } +q_f  \sqrt{H^2 q_f - 1} -3 \frac{q_f^2 }{2 \beta_1} , \\
&\text{Im}\left(\frac{S _{-1 , -1}}{V_3}\right)= \frac{ \sqrt{H^2 q_f-1} }{H^2} \left(5 H^2 q_f+ 3 \beta _1^2 H^4 -2\right) 
-\frac{ H^4 }{2} \beta _1^3 - 6 \beta _1 \left(H^2 q_f-1\right)- 3 \frac{q_f^2}{2 \beta _1} \,,
\end{align*}
with the standard phase dependence of the HH saddle point $ N_{-+} $ recovered only if by chance $\beta_1 = \frac{q_f}{2  \sqrt{H^2 q_f - 1}}$ (this condition can only be fulfilled for one particular value of the final size of the universe, i.e. it will generically not hold). As a consequence, the wave function is in general not real. However, the wave function will become real in the limit of large $ q_1 $, as the symmetry between the saddles is restored. Moreover, both of the relevant saddle point geometries behave classically in a WKB sense, since they have rapidly evolving phases as the final size of the universe is increased, and (to leading order) constant weighting/amplitudes. In the actual universe we would expect the interference to disappear as soon as perturbations are generated, since these effectively lead to the decoherence of the two background spacetimes \cite{Joos:1986iw, Kiefer:1987ft, Halliwell:1989vw}.

 
In concluding this section, let us say a few more words about interpretation. As mentioned before, when $\beta_0$ is negative and imaginary, the Robin boundary term in the action can be interpreted as an initial coherent state with imaginary momentum $ 3 V_3 \alpha_0 $ (see also~\cite{DiTucci:2019xcr, Vilenkin:2018dch}), 
\begin{align}
\Psi &= \int dN \, {\cal D}q \, dq_0  \, e^{iS/\hbar} \, \Psi_0\,, \quad
\Psi_0 \propto e^{ i 3V_3 \alpha_ 0 q_0 - 3 V_3 \frac{q_0^2}{2 |\beta_0|}} \, .
\end{align}
One can think of the original formulation of the no-boundary wavefunction as having as initial state a delta function centred at zero size $\delta(q_0)$, inevitably leading to unsuppressed fluctuations. We are dealing here instead with a Gaussian state peaked around zero, with a spread whose range is determined solely by the cosmological constant. This can be considered as a successful, minor modification of the original no-boundary wavefunction, where the HH geometry is actually dominant and singular geometries are avoided. Note however that the original state already encodes fluctuations of the spacetime geometry, i.e. the universe does not arise out of pure nothingness, but rather out of spacetime fluctuations. In some sense this is to be expected on grounds of the uncertainty principle, when applied to the spacetime geometry.

The (canonical) Robin boundary term in the action can equally well be interpreted as arising from a different state, 
\begin{equation}
\Phi_0 \propto e^{i 3 V_3 \gamma_0 q_0 - 3 V_3 \frac{(q_0 - q_i)^2}{2 |\beta_0|}} \, ,
\end{equation}
if $\gamma_ 0 = - \frac{q_i}{\beta_0} + \alpha_ 0  = - \frac{q_i}{\beta_0}  -  i  $.  This is possible because $\alpha_0$ and $ \beta_0 $ are imaginary in this implementation of the no-boundary wavefunction, and the imaginary piece $- \frac{q_i}{\beta_0}$ can be absorbed into a redefinition of the momentum. 
In particular, if $ \bar{q}_i=-\gamma_0 \, \beta_0 $ one can rewrite the state as 
\begin{equation}\label{key}
\Phi_0 \propto e^{ - 3 V_3 \frac{(q_0 - \bar{q}_i)^2}{2 |\beta_0|}} \, ,
\end{equation}
where now the mean momentum is zero, and the mean size is $\bar{q_i} =i\beta_0 \approx \frac{2}{H^2} \neq 0.$ In this case the central values of the scale factor and momentum are real, yet we still obtain complex saddle points of the path integral because the real values chosen are classically impossible (classically, the momentum is only zero at the waist of the de Sitter hyperboloid, where $q=1/H^2,$ while here we would demand the momentum to be zero at a larger scale factor value). This rewriting reinforces the point that we can no longer interpret this result as tunneling from nothing, even if the dominant saddle points are the complex Hartle-Hawking saddles.

Finally, it may be useful to contrast the present result with the results of the recent paper \cite{DiTucci:2019xcr} expounding a quantum incompleteness problem of inflation. In that paper, it was shown that a universe that starts out (i.e. evolves from zero size) in an inflationary phase generally does not acquire the Bunch-Davies vacuum for the fluctuations, but rather the fluctuations are unstable. A possible resolution was to put in an appropriate initial state, so that effectively the universe already starts at a large enough size. In that paper,
the initial state had a Lorentzian momentum, and in fact the relation between momentum and size was such that it could correspond to a classical solution. As a consequence the relevant saddle point was real. We cannot do the same here, if we want to obtain the Hartle-Hawking wavefunction, since this requires complex saddle points (one would otherwise not be able to smoothly ``round off'' the saddle point geometry). This is another way to see that the initial state must have momentum and size in a classically impossible configuration. This ``quantum'' initial state then provides a possible resolution of the incompleteness problem.



\section{Discussion}\label{sec:discussion}

The no-boundary wavefunction remains a leading theory of the initial conditions of the universe, as it has the potential to describe the emergence of spacetime, the early approach to classicality, the Bunch-Davies vacuum of perturbations and to provide initial conditions for geometry and scalar field evolution \cite{Hartle:2008ng}. However, in some sense it still remains an answer waiting for the appropriate question. One possibility is to simply define the no-boundary wavefunction by the properties just listed, and an approach in that general spirit was put forward recently in \cite{Halliwell:2018ejl,deAlwis:2018sec}. Another possibility, which many people have pursued and to which we are adhering here, is to find a suitable path integral definition of the no-boundary idea. After all, it is in this context that the idea was originally formulated \cite{Hawking:1981gb}.

In the 1980s it was thought that Euclidean quantum gravity might offer the best framework for quantum cosmological questions \cite{Gibbons:1994cg}. This eventually led people to explore variants of the integration contour for the lapse function, generalising the more obvious Lorentzian and Euclidean choices \cite{Halliwell:1988ik}. In the present paper we have (mostly) stuck to the Lorentzian contour for the lapse function, but have explored another aspect of the path integral formulation, namely the imposition of more general boundary conditions, both on the ``no-boundary'' hypersurface and at late times.  

We have focussed on three potential definitions of the no-boundary wavefunction. The way in which boundary conditions are imposed in general relativity is by specifying suitable boundary terms in the action. Then, under variation, a specific condition on the boundary values of the fields is implied. In the spirit of the no-boundary idea, we have first explored the case where one does not add any boundary term to the Einstein-Hilbert action. Despite the absence of a boundary term, boundary conditions are nevertheless implied for the fields. In particular we have reviewed how the absence of a boundary term implies a Neumann condition for the scale factor of the universe, allowing us to  impose a suitable Euclidean value for the derivative of the scale factor at the initial ``time''. This Euclidean derivative encodes the idea that the geometry must contain a Euclidean (or approximately Euclidean) section, thus allowing the spacetime to be closed off smoothly. Somewhat surprisingly, the absence of a boundary term imposes different initial conditions, namely Dirichlet conditions, for anisotropies in the Bianchi IX metric. Regularity of the saddle point solutions then requires that one set the initial anisotropy values to zero, whereafter they may grow towards the final hypersurface. With final Dirichlet conditions also being imposed on the final hypersurface for all fields, this ``no boundary term'' proposal reproduces the no-boundary wavefunction. 

In this construction, there is just one potential blemish: in deforming the contour of integration from the Lorentzian contour to the steepest descent paths of the integrand, one must pass through geometries that contain at least one singularity. This may leave one vulnerable to potentially large corrections upon inclusion of expected higher-derivative quantum correction terms in the gravitational action. (The contentious point is however whether or not one should be worried about this: after all, in the path integral one is summing over all paths, the majority of which will presumably contain numerous singularities.)
 We have shown that one may evade this potential problem by changing not the conditions at the initial hypersurface, but rather at the final one: instead of imposing a Dirichlet condition, one may impose a Robin condition. This is not simply of mathematical interest, but the main advantage is in fact that such conditions are physically sensible, as they allow one to specify the Hubble rate on the final hypersurface. One might argue that this is in any case more realistic, since the flatness of the presently observed universe does not allow us to measure the size of the universe whereas we can obtain the Hubble rate rather directly from redshift measurements. The interesting consequence is that the imposition of covariant Robin conditions pushes the singular geometries out of the way of the steepest descent contours. There is one price to pay however: the Euclidean momentum imposed on the initial hypersurface requires the contour of integration of the lapse function to be shifted away from the Lorentzian contour by a constant imaginary offset. It remains an open question whether this offset of the contour may lead to other consequences. We leave this question for future work. Another open question is whether one can construct other classes of quantum cosmological amplitudes (not necessarily in the no-boundary context) using covariant Robin conditions. In preliminary investigations of this question we have found that the presence of a square root in the covariant boundary action \eqref{actionRob} often leads to branch cuts in the action for the lapse. At present it is difficult to determine whether the appearance of branch cuts is only an artefact of the order of integration: it is entirely conceivable that the full infinite dimensional thimble involving both the integration over the lapse and over the scale factor encounters no issue with branch cuts. This is, however, a hard technical challenge which we must leave for future work.
 
There exists a different way of imposing Robin boundary conditions, which consists of treating the canonically normalised scale factor (squared) $q$ as the fundamental variable, and simply adding Robin boundary terms in a canonical manner. Although this imposition of Robin conditions is not covariant, it has an intuitive interpretation as imposing initial or final (in general complexified) Gaussian states. For a range of such states, we have shown that the path integral indeed reproduces the Hartle-Hawking wavefunction, and moreover the calculation is 
robust in that singular geometries may be avoided along the steepest descent contours, and in between these and the original Lorentzian contour of integration for the lapse function. An open question however remains as to whether a covariant version of these boundary terms may be found. 

The initial (or final) Gaussian states imply a shared uncertainty between the size of the universe and the expansion rate. In a sense, in this particular implementation of the no-boundary proposal the wavefunction may not describe creation from nothing, but rather creation from a coherent state of spacetime fluctuations. Indeed, the uncertainty principle, when applied to the geometry of spacetime, may not be compatible with true nothingness, which would demand knowing with certainty that both the size and the momentum of the universe were zero. But if there exists a ``primordial'' state of spacetime fluctuations, then one may wonder how this state originated and what else could  arise out of it.

Taking a step back from such speculative interpretive questions, the main result of the present work is that all current working models of the no-boundary wavefunction force a departure from the notion of a sum over compact metrics as the basic definition. Indeed, in order to avoid problems with unstable fluctuations, we have clarified that the imposition of an appropriate Euclidean momentum on the initial boundary guarantees the existence of smooth no-boundary saddle point geometries with stable fluctuations, as originally envisioned. However, off-shell these new definitions all involve sums over universes of various initial sizes, thereby offering the prospect that the physical interpretation of the no-boundary wavefunction may require further exploration and may end up being even richer than currently assumed.


\acknowledgments

We would like to thank Neil Turok and Sebastian Zell for discussions. ADT and JLL gratefully acknowledge the support of the European Research Council in the form of the ERC Consolidator Grant CoG 772295 ``Qosmology''. LS acknowledges that research at Perimeter Institute is supported by the Government of Canada through Industry Canada and by the Province of Ontario through the Ministry of Research and Innovation.


\appendix

\section{The path integral satisfies the WdW equation }\label{WdWproof}

We show in this section that the path integral with initial canonical Robin condition satisfies the homogeneous or inhomogeneous Wheeler--DeWitt (WdW) equation depending on the defining integration contour. Let us consider the normalised initial state 
\begin{equation}
\Psi(q_0) = \sqrt{\frac{6 V_3}{ \pi i \beta_0}} \, e^{i 3 V_3 \alpha_0 q_0 + i  3 V_3 \frac{q_0^2}{ 2  \beta_0}}\,.
\end{equation}
Note that $\Psi(q_0)$ is a coherent state for $\beta_0 = - i |\beta_0|$. 
The path integral with Robin boundary conditions can be written as follows 
\begin{equation}
\Psi(q_1) = \int dN \,  d q_0 \,  \delta q \, e^{ i S_D} \,  \Psi(q_0) \, ,
\end{equation} 
where $S_D$ is the appropriate action for the Dirichlet problem for gravity i.e. the Einstein-Hilbert bulk term plus the Gibbons-Hawking-York boundary term,
\begin{equation}
S_D= S_{EH}+S_{GHY} = 3 V_3 \int_0^1  \left[ - \frac{\dot{q}^2}{4 N} + N (1 - H^2 q)\right] \,.
\end{equation}
The integral can be written as follows
\begin{equation}
\Psi(q_1) = \sqrt{\frac{6 V_3}{ \pi i \beta}}  dN \,  d q_0 \, \int \delta q \, e^{i S_0}\,,
\end{equation}
where
\begin{equation}
\frac{S_0}{3 V_3} =  \int_0^1  [ - \frac{\dot{q}^2}{4 N} + N (1 - H^2 q)] + \alpha_0 q_0 + \frac{q_0^2}{2 \beta_0}
\end{equation}
is the total action, including the initial canonical boundary term.
The functional integral over $q$ gives
\begin{equation}
\Psi(q_1)  =\sqrt{\frac{6 V_3}{ \pi i \beta_0}} \int d N \,  d q_0 \sqrt{\frac{i 3 V_3 }{4 N}} e^{i S}\,,
\end{equation}
with 
\begin{equation}
\frac{S}{3 V_3} = \frac{1}{3} \left(\frac{H^4 N^3}{4} - \frac{3 (q_1 - q_0)^2}{4 N} + 3 N \Bigl( 1 - \frac{H^2}{2}(q_1 + q_0)\Bigr) \right) + \alpha_0 q_0 + \frac{q_0^2}{2 \beta_0}\,.
\end{equation}
To implement the no-boundary proposal, we need to fix $\alpha_0 = -  i  $. Integrating over $ q_0 $, we find
\begin{equation}
\Psi(q_1 ) =  - i \sqrt{\frac{3 V_3}{2 i }}\int \frac{d N}{\sqrt{2 N  -  \beta_0 }} e^{i \overline{S}} \,.
\end{equation}
The action $ \overline{S} = S(q_1 , \overline{q}_0)$ is evaluated at the saddle point $\overline{q}_0 = \frac{\beta_0 (q_1 - N(2 i + H^2 N ))}{\beta_0  - 2 N}$,
\begin{equation}
\begin{split}
\frac{\overline{S}}{3 V_3} = \frac{1}{6 \left(2 N - \beta_0 \right)} \Bigl(&H^4 N^4 - 2 \beta_0  H^4 N^3 + 6 N^2 (2 - H^2 ( i \beta_0   +   q_1)) + \\
& + 6 \beta_0  H^2 q_1 N + 3 q_1 (2 i \beta_0  -  q_1) \Bigr) \, .
\end{split}
\end{equation}
To evaluate the WdW equation, we need to compute
\begin{align}
\frac{\partial^2 \Psi(q_1)}{\partial q_1^2} & =   - i \sqrt{\frac{3 V_3}{2 i }} \int \frac{dN }{\sqrt{2 N -  \beta_0 }} \Bigl[i \frac{\partial^2 \overline{S}}{\partial q_1^2} - \Bigl( \frac{\partial \overline{S}}{\partial q_1}\Bigr)^2 \Bigr] e^{i \overline{S}} \, . 
\end{align}
We find
\begin{equation}
\begin{split}
 \int \frac{dN }{\sqrt{2 N -\beta_0}} i   e^{i \overline{S}} \frac{\partial^2 \overline{S}}{\partial q_1^2}  = &- 3 i  V_3 \int \frac{d N}{[2 N -\beta_0]^{3/2}} \, e^{i \overline{S}} = \\
 =&- 3 i V_3  \Bigl[ \int \frac{d N}{\sqrt{2 N -\beta_0}} \, i \,  \frac{\partial \overline{S}}{\partial N} \, e^{i \overline{S}} - 
 \Bigl( \frac{e^{i \overline{S}}}{\sqrt{2 N -\beta_0}}\Bigr)\Bigr|_{boundary} \Bigr]\,.
 \end{split}
\end{equation}
Therefore 
\begin{align}
\frac{\partial^2 \Psi(q_1)}{\partial q_1^2} =  &- i \sqrt{\frac{3 V_3}{2 i }}  \int \frac{dN}{\sqrt{2 N -\beta_0}} e^{i \overline{S}} \Bigl[ - \Bigl(\frac{\partial \overline{S}}{\partial q_1} \Bigr)^2 + 3 V_3  \frac{\partial \overline{S}}{\partial N}\Bigr] + \sqrt{\frac{(3 V_3)^3}{2 i }}  \Bigl( \frac{e^{i \overline{S}}}{\sqrt{2 N -\beta_0}}\Bigr)\Bigr|_{boundary} \nonumber \\
= & \, i \sqrt{\frac{(3 V_3)^5}{2 i }} \int \frac{dN}{\sqrt{2 N -\beta_0}}  \Bigl[(H^2 q_1 - 1) \, e^{i \overline{S}} \Bigl]+ \sqrt{\frac{(3 V_3)^3}{2 i }} \Bigl(\frac{e^{i \overline{S}}}{\sqrt{2 N -\beta_0}} \Bigr) \Bigr|_{boundary}
\end{align}
If we consider an integration contour which runs from $N \rightarrow - \infty$ to $N \rightarrow + \infty$ along the real line $\Psi (q_1 )$ satisfies the WdW equation. Indeed, in this case the boundary term vanishes,
\begin{equation}
\lim_{N \rightarrow \pm \infty} \frac{e^{i \overline{S}}}{\sqrt{2 N -  \beta_0 }} = 0 \, . 
\end{equation}
We thus obtain that $\frac{\partial^2 \Psi(q_1)}{\partial q_1^2}  =   - 9 V_3^2 (H^2 q_1 - 1) \Psi(q_1)$.  

The other possibility is to take the contour to run from the singularity at $N^* = \frac{\beta_0}{2}$ to $N \rightarrow + \infty$.  In order to calculate the boundary term, notice that for $N \approx N^*$ the action diverges as
\begin{equation}
\frac{\overline{S}}{3 V_3} \approx - \frac{1}{32 \left(2 N -  \beta_0 \right)} \left(4 q_1 -  \beta_0  (4 i + \beta_0 H^2)\right)^2 \,.
\end{equation}
It was shown in~\cite{DiTucci:2019dji} and in Section~\ref{sec:canonical} that the relevant case for the no-boundary proposal is when $\beta_0$ takes negative imaginary values, $\beta_0 = - i |\beta_0|$. In this case the singularity $N^* $ lies on the negative imaginary axis and the thimble approaches it along the axis, so that $N \sim i \, n$ as $N \rightarrow N^*$. 
The boundary term at the singularity is then proportional to a Dirac delta function
\begin{equation}
\begin{split}
\lim_{N \rightarrow N^*}  \frac{e^{i \overline{S}}}{\sqrt{2 N - \beta_0 }} & = \lim_{x \rightarrow 0 } \sqrt{2 \pi i } \frac{1}{\sqrt{2 \pi i  x}} \, e^{- \frac{i 3 V_3}{4 x} \left(q_1 -\frac{\beta_0}{4}(4 i +  \beta_0  H^2)\right)^2} \\ 
&= \sqrt{\frac{4 \pi}{3  i V_3 }} \, \delta ( q_1 - \frac{\beta_0}{4} (4 i + \beta_0 H^2)) \, ,
\end{split}
\end{equation} 
where $x = N - N^* $ is purely imaginary. Therefore
\begin{equation}
\frac{\partial^2 \Psi(q_1)}{\partial q_1^2}  =   - 9 V_3^2(H^2 q_1 - 1) \Psi(q_1) - 3 V_3 i  \sqrt{2 \pi}  \,  \delta ( q_1 - \frac{\beta_0}{4} (4 i + \beta_0 H^2))\,,
\end{equation}
i.e. $\Psi$ satisfies the inhomogeneous WdW equation.

\bibliographystyle{utphys}
\bibliography{Ref}

\end{document}